\documentclass[prl,aps,twocolumn,tightenlines,floatfix,nofootinbib]{revtex4}

\usepackage{epsf}    
\usepackage{graphicx}
\usepackage{epstopdf}

\newcommand{\be}{\begin{equation}}
\newcommand{\ee}{\end{equation}}
\newcommand{\bear}{\begin{eqnarray}}
\newcommand{\eear}{\end{eqnarray}}

\newcommand{\bA}{\mathbf{A}}
\newcommand{\bB}{\mathbf{B}}
\newcommand{\bnabla}{\mathbf{\nabla}}
\newcommand{\bz}{\mathbf{z}}
\newcommand{\bphi}{\mathbf{\varphi}}
\newcommand{\bJ}{\mathbf{J}}

\newcommand{\rp}{{\rm p}}

\newcommand{\rX}{{\rm X}}

\newcommand{\cA}{{\cal A}}
\newcommand{\cC}{{\cal C}}

\newcommand{\cE}{{\cal E}}

\newcommand{\cN}{{\cal N}}

\begin{document}

\title{Gravitational waves from color-magnetic `mountains' in neutron stars}

\author{K. Glampedakis$^{1,2}$, D.I. Jones$^3$ and  L. Samuelsson$^{4,5}$ }
\affiliation{$^1$Departamento de F\'isica, Universidad de Murcia, E-30100 Murcia, Spain \\
$^2$Theoretical Astrophysics, University of T\"ubingen, D-72076 T\"ubingen, Germany\\
$^3$School of Mathematics, University of Southampton, Southampton SO17 1BJ, UK\\
$^4$Department of Physics, Ume\aa\ University, SE-901 87 Ume\aa, Sweden \\
$^5$Nordita, Roslagstullsbacken 23, SE-106 91 Stockholm, Sweden}

\begin{abstract}
Neutron stars may harbour the true ground state of matter in the form of strange quark matter.
If present, this type of matter is expected to be a color superconductor, a consequence of quark pairing with 
respect to the color/flavor degrees of freedom. The stellar magnetic field threading the
quark core becomes a color-magnetic admixture and, in the event that superconductivity is of type II,
leads to the formation of color-magnetic vortices. In this Letter we show that the volume-averaged color-magnetic
vortex tension force should naturally lead to a significant degree of non-axisymmetry in systems like radio pulsars.
We show that gravitational radiation from such color-magnetic `mountains' in young pulsars like the Crab and Vela 
could be observable by the future Einstein Telescope, thus becoming a probe of paired quark matter in neutron stars. 
The detectability threshold can be pushed up toward the sensitivity level of Advanced LIGO if we invoke an interior magnetic
field about a factor ten stronger than the surface polar field.

\end{abstract}

\maketitle


{\em Introduction}.--- Neutron stars might actually be quark stars in the event that stable strange quark matter
is realised in the high density/pressure environment of these objects~\cite{alcock86}. Confirming (or ruling out)
the existence of this type of matter ranks among the ultimate goals of neutron star astrophysics. But this will not
easily come about; quark stars may masquerade as neutron stars with respect to the bulk properties of mass and 
radius~\cite{alford05}. The situation is exacerbated if the star is a hybrid, with a quark inner core cloaked by a mantle
of hadronic matter \cite{glenbook}. Probing, for instance, the physics of the crust may reveal nothing about a quark core.
Similarly, if stellar pulsation modes were to be observed, their properties could be controlled by the hadronic outer core and crust (e.g. this is likely the case for the $r$-mode instability~\cite{madsen,nils}). Neutron star cooling, a potentially powerful probe of stellar structure, could also prove to be largely insensitive to the physics of an inner quark core~\cite{review08}.

If present, quark matter is likely to be in its stablest form as a condensate of paired quarks. Among the various
possible states, the two most intensively studied are~\cite{review08}: (i) the color-flavor-locked (CFL) 
phase where all three quark species $(u,d,s)$ are paired (ii) the two-flavor phase (2SC) where only the $u$ and $d$ quarks pair.
In both phases the system is a `color superconductor', an exotic variant of ordinary electric superconductivity.  

This key property of quark matter is the subject of the present Letter. The large scale magnetic field present in
neutron stars can penetrate the CFL/2SC region, inducing `color-magnetic' stresses which deform the stellar core, much like
the magnetic deformation in ordinary neutron stars~\cite{chandra}. As we discuss in detail below, in the event that the system 
behaves as a type II superconductor the deformation is greatly amplified by the formation of color-magnetic vortices, each one
carrying an energy $\sim 10^3$ times higher than the energy of ordinary protonic vortices (a result of the stronger Meissner screening associated with color superconductivity). 

If systems like radio pulsars do contain a color-superconducting quark core, threaded by a large-scale magnetic field misaligned with the rotation axis, they should display a significant non-axisymmetric deformation in the form of a color-magnetic `mountain'. As a result of this, these systems will emit a continuous gravitational wave signal, at a level much stronger than that due to magnetic field deformation in their hadronic counterparts.  Our results on the detectability of such signals suggest that young pulsars -- most notably the Crab -- could be observed by future gravitational wave detectors like the Einstein Telescope (ET) and possibly even Advanced LIGO.


{\em Color superconductivity in neutron stars}. --- A magnetic field threading a region of 
color-superconducting quark matter is `rotated' in the space of internal gauge degrees of freedom, and becomes an admixture of
`color-magnetic' fields (see Ref.~\cite{review00} for a review). Among the different possible combinations, 
the following mixed gauge field $\bA^\rX$ is associated with a massive propagator and superconducting Meissner 
screening~\cite{review00}
\be
\bA^\rX = \sin \chi \bA + \cos\chi \bA^8 ,
\label{Ax}
\ee
where $\bA$ and $\bA^8$ are the electromagnetic and (one of) the gluonic vector potentials, respectively.
The `mixing' angle $\chi$ is (essentially) the electromagnetic to gluonic coupling constant ratio 
(expressed in $\hbar=c=1$ units)
\be
\sin\chi = \left \{\begin{array}{ll}
                ( 1 + 3g^2/4e^2 )^{-1/2} &\quad \mbox{(CFL phase),} \\
                             & \\
                ( 1+ 3g^2/e^2 )^{-1/2}  &\quad  \mbox{(2SC phase).}
             \end{array} 
     \right.
\label{angle} 
\ee
The slightly different numerical factors in these expressions reflect the different quark pairing in the two phases. 

The disproportionate coupling strengths, $e^2 \approx 4\pi/137$ and $g \approx 3.5$ in natural units, result in 
$\chi \ll 1$ which in turn implies that the field $\bA^\rX$ is mostly gluonic. The bulk electromagnetic field is 
insensitive to the quark color pairing/superconductivity and the system behaves paramagnetically with respect to
it~\cite{review00,review08}.

The potential (\ref{Ax}) is associated with a color-magnetic field $\bB^\rX = \bnabla \times \bA^\rX$
which obeys an Amp\'ere-type law $\bnabla \times \bB^\rX = 4\pi \bJ^\rX /c$
where $\bJ^\rX$ is the combined color-electric current~\cite{iida05}. 
In the context of laboratory superfluidity/superconductivity, the mass/electric current
makes direct contact with the macroscopic quantum state of the paired system. The same is true for the
more exotic quark color superconductor. The current has a form $\bJ^\rX = k_1 \nabla \theta + k_2 \bA^\rX$
where $k_1, k_2, \theta$ depend on the superconducting phase (see~\cite{IB02,iida05} for the explicit expressions). 
The (spatially varying) $\theta$ is the so-called order parameter, that is, the phase of the macroscopic 
wavefunction.

Color superconductivity displays the familiar type I/II dichotomy. Which type prevails in the
conditions met in neutron stars is largely determined by the pairing gap energy to quark chemical potential 
ratio $\Delta/\mu_{\rm q}$~\cite{AS10}. The formal criterion for determining the type of superconductivity is the same as
in electromagnetic systems: type II is the energetically favorable state provided the Ginzburg-Landau 
parameter $\kappa = \Lambda/\xi > 1/\sqrt{2} $, where $\Lambda$ and $\xi$ are, respectively, the penetration and coherence lengths.
The situation is blurred by the fact that the underlying theory is actually an approximation near the
critical temperature $T_c$ for the onset of superconductivity. Extrapolating to the temperature regime $T \ll T_c$ (appropriate for astrophysical neutron stars) for the 2SC phase, Alford \& Sedrakian~\cite{AS10} estimate $\kappa \approx 11 \Delta/\mu_{\rm q}$ which predicts a type II state for $\Delta \gtrsim 25\,\mbox{MeV} $ (assuming a typical quark chemical potential $\mu_{\rm q} = 400\,\mbox{MeV}$). The theoretically predicted gaps lie in the range $\Delta \sim 20-100\,\mbox{MeV}$ \cite{review08} and therefore most of them satisfy this condition. The analysis for the CFL state is equally uncertain. The results of \cite{giannakis} suggest $\kappa \approx 9 \Delta /\mu_{\rm q}$  which again predicts type II superconductivity for `typical' pairing gaps.  

Without any additional microphysics input we hereafter assume that CFL/2SC color superconductivity is of 
type II and that the $\bB^\rX$ field forms a stable Abrikosov lattice of color-magnetic vortices. 
The formation of vortices manifests in the form of singularities in the order parameter $\theta$ at the location of each 
vortex. Skipping the details that are not essential for our present discussion, the end-product is the formulation of the 
following quantization condition for an isolated vortex (see \cite{IB02}):
\be
\bA^\rX + \frac{4\pi\Lambda^2_\star}{c}  \bJ^\rX  = \frac{\phi_\star}{2\pi r}  \hat{\bphi} ,
\label{quant1}
\ee
where the asterisk stands for $\star = \{\mathrm{2SC},\mathrm{CFL} \}$ and $\phi_\star$ is the color-magnetic 
flux quantum (see below). The vortex singularity is located at $r=0$ with the vortex itself identified 
with z-axis in cylindrical coordinates $\{r,\varphi,z\}$.

The condition (\ref{quant1}) applies on the {\it mesoscopic} neighborhood of a single vortex, where the matter  
can be assumed to be uniform (this entails $\Lambda_\star \approx \mbox{constant}$). Taking the curl of (\ref{quant1}) 
leads to the London equation \cite{IB02,iida05}
\be
\nabla^2 \bB^\rX - \frac{1}{\Lambda_\star^2} \bB^\rX = -\frac{\phi_\star}{\Lambda_\star^2} \delta (\bf{r}) \hat{\bz}
\label{LondonX} .
\ee
This equation encapsulates the Meissner-screening property of the $\bB^\rX$ field. 


{\em The color-magnetic vortex energy}.--- Using the imposed symmetries of our single vortex system 
the above London equation reduces to an ODE with the familiar Bessel-function solution~\cite{tilley},
\be
\bB^\rX = \frac{\phi_\star}{2\pi \Lambda_\star^2} K_0 \left (\frac{r}{\Lambda_\star} \right ) \hat{\bz} .
\label{Bxsol}
\ee
This field is associated with the flux $\int {\bf dS} \cdot \bB^\rX = \phi_\star$. 
The total energy per unit vortex length consists of contributions from the color-magnetic field and current,
\be
\cE_\rX =   \int dS \frac{(B^\rX)^2}{8\pi}  + \frac{\pi\Lambda^2_\star}{4 c^2} \int dS (J^\rX)^2 .
\label{energ}
\ee
It is straightforward to compute the integrals using (\ref{Bxsol}).
If we furthermore assume a strong type II superconductivity, i.e.  $\xi_\star \ll \Lambda_\star $, the
final result is  
\be
\cE_\rX \approx \frac{\phi_\star^2}{16\pi^2 \Lambda^2_\star} \log \left ( \frac{\Lambda_\star}{\xi_\star} \right ) .
\label{Ecv}
\ee
Not surprisingly, this energy has a functional form identical to that of fluxtubes in a proton superconductor~\cite{mendell91}. 

We now need to input numerical values for $\Lambda_\star$ and $\phi_\star$. These are given in \cite{IB02};
the penetration length is
\be
\Lambda_\star \approx  10^{-13} \zeta_\star \left ( 1/g_\star \right ) \mu_{q400}^{-1}\,~ \mbox{cm}
\ee
where $\mu_{\rm q400} = \mu_{\rm q}/400\,\mbox{MeV}$, $\zeta_{\rm CFL} =1.1$, $\zeta_{\rm 2SC} =1.9$,   
\be
g_{\rm 2SC} = \frac{1}{3} \left ( 3 g^2 + e^2  \right )^{1/2}, \quad
g_{\rm CFL} = \frac{1}{6} \left ( 3 g^2 + 4 e^2  \right )^{1/2} .
\ee
Note that these coupling combinations are to be evaluated using natural units. For the flux quantums we have
\be
\phi_\star = hc/g_\star =  6 \sin\chi \phi_0 ,
\label{quantum}
\ee
where $\phi_0 = hc/2e \approx 2\times 10^{-7}\, \mbox{G}/\mbox{cm}^2 $.

Using these estimates and making the usual approximation $ (1/2) \log (\Lambda_\star/\xi_\star) \approx 1$ \cite{tilley},
we arrive at our final results for the color-magnetic vortex energy:
\bear
&& \cE_\rX^{\rm 2SC} \approx 4.9 \times 10^9 \mu_{\rm q400}^2\,~ \mbox{ergs}/\mbox{cm} .
\\
\nonumber \\
&& \cE_\rX^{\rm CFL} \approx 1.5\times 10^{10} \mu_{\rm q400}^2\,~ \mbox{ergs}/\mbox{cm} .
\eear
Most noteworthy, the vortices of both CFL and 2SC phases carry much more energy than the energy $\cE_\rp \approx 10^7\,\mbox{ergs}/\mbox{cm}$ of protonic fluxtubes~\cite{mendell91}.


{\em The stellar deformation}.---We can estimate the ellipticity due to the color-magnetic stresses in the quark core using the ratio of vortex to gravitational binding energy:
\be
\label{eq:epsilon_X}
\epsilon_\rX \approx \frac{\bar \cN_\rX \cE_\rX V_{\rm q}}{3G M^2/(4R)} ,
\ee
where $\bar \cN_\rX$ is the vortex area density averaged over the quark core, $V_{\rm q}$ is the volume of the core, and we have used the gravitational binding energy appropriate to an $n=1$ polytrope [hereafter, we scale our results to a mass $M=1.4 M_\odot$, a radius $R = 11.7$ km, and a moment of inertia $I = 10^{45}$ g cm$^2$, consistent for an $n=1$ polytrope].
It should be pointed out that (\ref{eq:epsilon_X}) is an approximation~\cite{chandra}; 
nevertheless, it is known to be quite accurate in reproducing rigorously computed ellipticities 
of magnetically deformed neutron stars (see~\cite{akgun} for a discussion of the superconducting case).

Following Ref.~\cite{AS10}, we can relate $\cN_\rX$ to the {\it macroscopic interior} magnetic field $B$:
\be 
\cN_\rX = \sin\chi B / \phi_\star = \cN_\rp /6 \approx 8.1 \times 10^{17} B_{12}\,~\mbox{cm}^{-2} ,
\ee
where $\cN_\rp = B/\phi_0$ is the density of protonic vortices and $B_{12}= B/10^{12}\,\mbox{G}$. 
Parametrizing the quark core as $V_{\rm q} = f V_{\rm star}$, we obtain
\be
\label{eq:epsilon_2SC}
\epsilon_\rX^{\rm 2SC} \approx 8.0 \times 10^{-8}\, f \bar B_{12} \mu_{\rm q400}^2 R_{11.7}^4 M_{1.4}^{-2} , 
\ee
where $\bar B_{12}$ is the volume-average of $B$ over the quark core. 
The CFL-phase ellipticity is similarly found to be:
\be
\label{eq:epsilon_CFL}
\epsilon_\rX^{\rm CFL} \approx 2.4 \times 10^{-7}\, f \bar B_{12} \mu_{\rm q400}^2 R_{11.7}^4 M_{1.4}^{-2} . 
\ee
For comparison, the ellipticity resulting from an ordinary protonic superconducting star is 
(setting $f=1$),
\be
\epsilon_\rp \approx 9.7 \times 10^{-10} \bar B_{12} R_{11.7}^4 M_{1.4}^{-2} .
\ee


{\em Gravitational wave detectability}.--- It is straightforward to use these results to estimate the level 
of gravitational wave emission from known pulsars, assuming they do indeed contain color superconducting 2SC or CFL cores.  

First we need to use the observed spin $\Omega$ and spin-down rate $\dot \Omega$ to infer the magnetic field strength.  Note that the volume-averaged field $\bar B$ is likely to exceed the dipole surface polar field $B_{\rm p}$ by a factor of a few; certainly this is the case for known magnetic equilibria, see e.g. \cite{ciolfi09,lj09}, where $\bar B \approx 2 B_{\rm p}$.  We therefore define $\bar B = \alpha B_{\rm p}$.  For the electromagnetic spin-down torque we will use the usual vacuum dipole result: 
$\dot J_{\rm EW} = (R^6/6c^3) B_{\rm p}^2 \Omega^3 \equiv \cA B_{\rm p}^2 \Omega^3$. For the gravitational wave spin-down, we have 
$\dot J_{\rm GW} = (32GI^2/5c^5) \epsilon_{\rm X}^2 \Omega^5$  \citep{st83}, which, when combined with (\ref{eq:epsilon_2SC}) and 
(\ref{eq:epsilon_CFL}), takes the form  $\dot J_{\rm GW} = \cC_{\rm X} \alpha^2 B_{\rm p}^2 \Omega^5$. Then the balance 
$\dot{J}_{\rm EW} + \dot{J}_{\rm GW} = I \dot{\Omega}$ leads to
\begin{equation}
\label{eq:B}
B_{\rm p} = \left[\frac{I\dot\Omega}{\cA\Omega^3 + \cC_{\rm X} \alpha^2 \Omega^5}\right]^{1/2} .
\end{equation}
Note that $\dot{J}_{\rm GW}$ in (\ref{eq:B}) is negligible in all cases apart from rapidly spinning systems 
with $\alpha \gtrsim 100$.

The field $\bar B = \alpha B_{\rm p}$ can then be substituted into equations (\ref{eq:epsilon_2SC}), (\ref{eq:epsilon_CFL}) to give the corresponding ellipticities, which can then be converted into gravitational wave amplitudes using equation (55) of Ref.~\cite{thorne87}:
\be
h_{\mathrm{gw}} = \frac{8G}{c^4} \left(\frac{2}{15}\right)^{1/2} \frac{\epsilon_{\rm X} I \Omega^2}{d} ,
\ee
for a star a distance $d$ from Earth.  

\begin{figure}
\centerline{\includegraphics[height=6cm,clip]{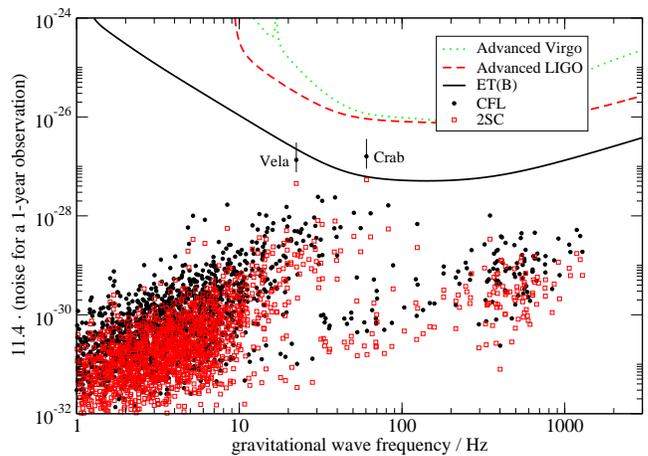}}
\caption{Estimated gravitational wave amplitudes for known pulsars,
  assuming they contain 2SC or CFL cores with color-magnetic
  ellipticities given by (\ref{eq:epsilon_2SC}) and
  (\ref{eq:epsilon_CFL}). We assumed a quark volume fraction $f =1/2$,
  a quark chemical potential of $\mu_{\rm q} = 400$ MeV, and a ratio
  of internal volume-averaged to polar field of $\alpha = \bar{B}/B_{\rm p}
  = 2$. For the Crab and Vela, the short vertical line segments show
  the range in wave amplitude as $\mu_{\rm q}$ varies over the
  interval $300$--$600$ MeV, assuming CFL cores.   All pulsar data was
  taken from the ATNF database \cite{ATNF}. The detector noisecurves were taken from \cite{aligo} and \cite{ET}.
\label{fig:paper_fig}}
\end{figure}

The results are shown in Figure~\ref{fig:paper_fig}, for both 2SC and CFL cores.  We have used sensible values for the various physical parameters, including a quark chemical potential $\mu_{\rm q} = 400$ MeV, a fiducial quark core volume fraction $f = 1/2$, an internal average magnetic field of $\alpha = \bar B /B_{\rm p} =2$.  Note that the calculated ellipticities, and therefore the gravitational wave amplitudes, are proportional to $\alpha f \mu_{\rm q}^2$, so that the results may scale up or down, depending upon the actual values of these parameters as realised in Nature. To capture some of this uncertainty, for the CFL estimates for the Crab and Vela pulsars, we have held all parameters fixed apart from $\mu_{\rm q}$, which we have varied over the interval $300$--$600$ MeV \cite{review08}, to give the two short line segments indicated in the Figure.

{\em Discussion}. --- Clearly, the Crab and Vela pulsars are the two most interesting candidates in terms of detection; both lie close to the ET detectability curve.  In the case of the Crab and a CFL core, our results suggest that, for the stellar parameters used here,  the Crab may be detectable by ET.  In the case of a 2SC core, detectability may be possible, but only if $\mu_{\rm q}$ lies in the upper part of the range quoted above.  In the case of the Vela and a CFL core, detection would be possible only if $\mu_{\rm q}$ lies close to the upper end of the quoted $\mu_{\rm q}$-range, while a 2SC core in Vela would not be detectable.  Clearly, the systems lie sufficiently close to the detectability curve that a confident statement regarding detectability cannot be made here; a more careful treatment is clearly needed.

There has been discussion as to the possibility of some pulsars harbouring much stronger fields than the spin-down estimates would suggest. In particular, the existence of `anomalous' braking indices $\ddot\Omega \Omega / \dot\Omega^2 <3$ has been taken as evidence of the gradual emergence of a buried strong magnetic field from the interior of some young pulsars by diffusive processes (see \cite{eetal11} and references therein). In fact, both Crab and Vela belong to this pulsar subgroup. Further weight has been added to this theory by observations of the rotation-powered pulsar PSR J1734, which seems to be evolving toward the magnetar portion of the pulsar $(\Omega, \dot \Omega)$ plane \cite{eetal11}.  This would correspond to the young pulsars having internal fields one to two orders of magnitude larger than their polar fields, i.e. $\alpha \sim 10$--$100$.  

In light of these suggestions, we give in Table~\ref{table:alphas} the \emph{minimum} $\alpha$ values required for detectability of the Crab and Vela in our model, assuming sensible values of all other parameters ($f=1/2$, $\mu_{\rm q} = 400$ MeV). 
Clearly, if the Crab and Vela do harbour strong internal fields, the corresponding CFL and 2SC emission would lie within range of ET, and possibly even Advanced LIGO.  We also note that the current $95\%$ confidence direct upper limit on gravitational wave emission from the Crab pulsar of $h^{95\%} = 2.0 \times 10^{-25}$, obtained using LIGO data \cite{known_pulsars_S5}, implies an upper limit of $\alpha \approx 250$ for this pulsar, assuming that it contains a CFL core. 
 
\begin{table}
\caption{Minimum value of $\alpha = \bar{B}/B_p$, the ratio of volume-averaged internal field to polar field, 
required for gravitational wave detection. We fix $f=1/2$ and $\mu_{\rm q} = 400$ MeV.}
\label{table:alphas}
\begin{tabular}{|c|c|c|c|c|} 
\hline
& \multicolumn{2}{c|}{ET} & \multicolumn{2}{c|}{Advanced LIGO} \\
\hline
& Crab & Vela & Crab & Vela \\
\hline
CFL & $\quad 0.78 \quad$ & $ \quad 3.3 \quad $ & $\quad 12 \quad $ & $46$ \\
2SC & $\quad 2.3 \quad $ & $9.9$ & $35$ & $137$ \\
\hline
\end{tabular}
\end{table}

In fact, for $\alpha \sim 100$, we find that $\sim 40$ pulsars are detectable by ET, including many millisecond pulsars.   Increasing $\alpha$ to $\sim 10^3$ pushes many of the millisecond systems close to their spin-down limits, leaving $\sim 30$ of them detectable by ET (such high $\alpha$ values could naturally materialise in recycled millisecond pulsars as a result of magnetic burial during the previous accretion-dominated stage of their lives). For these systems, the gravitational wave torque would then dominate the electromagnetic one, and the conventional electromagnetic-only calculation of $B_{\rm p}$ would significantly overestimate the actual polar field strength (see equation (\ref{eq:B})).   

Failure of ET to detect gravitational waves from the Crab or Vela does {\it not} rule out the existence of color superconducting quark cores in these (or any other) neutron  stars. It may be that the relevant parameters lie at the low end of the possible ranges, i.e. $\mu_{\rm q}\approx 300$ MeV, and/or the core occupies only a small fraction of the total stellar volume.  Alternatively, it may be that the internal field arranges itself in such a way as to give a smaller ellipticity than our estimate of equation (\ref{eq:epsilon_X}) would suggest. However, this would require a rather special geometry, not seen in magnetic fields studied in `normal' neutron stars (see e.g. \cite{lj09}).  Another possibility is that the vortex arrays discussed here simply aren't stable, in other words type II color-superconductivity
is not realised, resulting in a lower ellipticity than equation (\ref{eq:epsilon_X}) would predict.

However, if the link between field strength and deformation is similar to the well understood problem in non-color superconducting stars, our estimates should be reliable.
In this sense the link between CFL/2SC cores and gravitational wave emission is relatively robust, and more direct than the link between exotic \emph{elastic} cores and gravitational wave emission, where, in addition to the existence of the exotic core, a mechanism to generate the non-axisymmetric strains that perturb the star away from the lower energy axisymmetric state must also be provided \cite{owen05, hetal07, lin07, ks09}.

A more accurate calculation of the gravitational wave emission would self-consistently solve for a self-gravitating color 
superconducting star, accounting for the multi-component nature of the stellar matter and the magnetic field geometry, and differentiating between the quark core and hadronic envelope (see \cite{hetal07} for discussion of this last point). 
Equally crucial is the need for a more accurate input from microphysics, e.g. vortex structure calculations for the
density/temperature regime appropriate to neutron stars and consideration of a larger variety of color superconducting 
states~\cite{review08}. While challenging, the proximity of candidate sources to the ET detectability
curve clearly motivates the development of the tools needed to do this, to better assess the ability of gravitational wave observations to provide this unique probe of the state of matter at the highest densities.

\acknowledgements
KG is supported by the Ram\'{o}n y Cajal Programme in Spain. DIJ is supported by the STFC in the UK. 
LS is supported by the ERC (Contract No.\ 204059-QPQV), and the Swedish Research Council (Contract No.\ 2007-4422).  
The authors also acknowledge support from COMPSTAR, an ESF Programme, and thank Mark Alford and Keith Riles for comments on an earlier draft of the paper.


\end{document}